\documentclass[10pt,letterpaper]{article}
\usepackage[top=0.85in,left=2.75in,footskip=0.75in,marginparwidth=2in]{geometry}
\usepackage{amsmath}
\usepackage{breqn}
\usepackage{upgreek}
\usepackage{xcolor}

\usepackage{graphicx}
\usepackage{epstopdf}

\usepackage[utf8]{inputenc}

\usepackage{cite}

\usepackage{nameref,hyperref}

\usepackage{microtype}
\DisableLigatures[f]{encoding = *, family = * }

\raggedright
\usepackage{changepage}

\usepackage[aboveskip=1pt,labelfont=bf,labelsep=period,singlelinecheck=off]{caption}

\makeatletter
\renewcommand{\@biblabel}[1]{\quad#1.}
\makeatother

\usepackage{lastpage,fancyhdr,graphicx}
\usepackage{epstopdf}
\fancyheadoffset[L]{2.25in}
\fancyfootoffset[L]{2.25in}

\usepackage{color}

\definecolor{Gray}{gray}{.25}

\usepackage{graphicx}

\usepackage{sidecap}

\usepackage{wrapfig}
\usepackage[pscoord]{eso-pic}
\usepackage[fulladjust]{marginnote}
\reversemarginpar

\usepackage{ragged2e}
\justifying\let\raggedright\justifying

\begin{document}

\vspace*{0.35in}

\begin{flushleft}
{\Large
\textbf\newline{Ultra-subwavelength focusing and giant magnetic-field enhancement in a low-loss one-way waveguide based on remanence}
}
\newline
\\
Jie Xu,\textsuperscript{1,2,3} Xiaohua Deng,\textsuperscript{2} Hang Zhang,\textsuperscript{5,6} Chiaho Wu,\textsuperscript{5} Martijn Wubs,\textsuperscript{3,4} Sanshui Xiao,\textsuperscript{3,4} and Linfang Shen\textsuperscript{2,5,*}
\\
\bigskip
\bf{1} College of Material Science and Engineering, Nanchang University, Nanchang 330031, China
\\
\bf{2} Institute of Space Science and Technology, Nanchang University, Nanchang 330031, China
\\
\bf{3} Department of Photonics Engineering, Technical University of Denmark, DK-2800 Kgs. Lyngby, Denmark\\
\bf{4} Center for Nanostructured Graphene, Technical University of Denmark, DK-2800 Kgs. Lyngby, Denmark \\
\bf{5} Department of Applied Physics, Zhejiang University of Technology, Hangzhou 310023, China\\
\bf{6} physzhang@zjut.edu.cn
\\
\bigskip
* lfshen@zjut.edu.cn

\end{flushleft}

\section*{Abstract}
The subwavelength focusing based on surface plasmon polaritons (SPP) has been widely explored in tapered metallic structures. However, the efficiency of energy localization is relatively weak, largely due to  high propagation loss and strong back reflection. Here, we propose a straight-tapered 3-dimensional (3D) one-way surface magnetoplasmon (SMP) waveguide with the ending surface of $\sim 10^{-4}\lambda_0 \times 10^{-4}\lambda_0$ to achieve energy focusing in the microwave regime.
Due to low propagation loss of SMP, we achieve huge magnetic field enhancement in such an ultra-subwavelength area, by five orders of magnitude. Instead of using an external static magnetic field, our proposed SMP waveguide relies on remanence, which is very convenient for operating practical 3D applications. These results show promising applications in magnetic-field enhancing or quenching fluorescence, luminescence or nonlinearity of 2D materials, novel scanning near-field microwave microscopy and energy storage.


\section{Introduction}
Similar to the chiral edge states founded in the quantum-Hall effect \cite{Prang:1}, one-way propagation of electromagnetic (EM) modes increasingly attracts attention in recent years\cite{Haldane:2,Skirlo:3,Lu:4,Khanikaev:5}. A surface magnetoplasmon (SMP) sustained at the interface of a magnetic-optical (MO)  and a dielectric material behaves as a one-way EM mode when applying  an external dc magnetic field on MO material to break the time-reversal symmetry\cite{Raghu:6,Wang:7,Ao:8,xu:9,Shen:10}. The one-way SMP was experimentally observed by Wang et al. in 2009 by using gyromagnetic yttrium-iron-garnet (YIG) in the microwave regime\cite{Wang:11}. Recently, it was reported that one-way SMPs can be used to build zero-dimensional\cite{Dutra:12} cavity, which breaks the time-bandwidth limit\cite{Tsakmakidis:13}, and the study by Mann et al. demonstrates that for overcoming the time-bandwidth limit, time-varying systems or nonlinearities are required\cite{Mann:14}. More recently, we proposed a high-quality subwavelength isolator based on remanence (We note that remanence has been widely used in industry\cite{Pozar:32}) and we theoretically investigated this one-way SMP\cite{You:15}. The one-way feature of the SMP mode provides a promising way to realizing light focusing, because it suppresses the reflections.

In traditional optics, light focusing suffers from the diffraction limit. It was reported that surface plasmon polaritons (SPP)\cite{Stockman:16,Yin:17,Steele:18,Desiatov:19,Choo:20,He:21} or spoof plasmon polaritons\cite{Maier:22,Zhang:23} or channel plasmon polaritons\cite{Volkov:24,Smith:25,Moreno:26,Sondergaard:27} can achieve subwavelength focusing, in which this limit is overcome. One of the main advantages of subwavelength focusing is EM field enhancement. Most of previous works focus on the enhanced electric field, instead of the magnetic field \cite{Nazir:28,Mirzaei:29} or \cite{Shen:10}. Moreover, the efficiency of the energy localization previously demonstrated is relatively low because of the strong reflection, substantial radiation loss, or both.

In this paper, we theoretically analyze and design a straight-tapered 3D low-loss one-way waveguide based on remanence to achieve subwavelength focusing. After carefully analyzing the propagation properties of the guided modes in both lossless and lossy conditions, we set up a tapered configuration with an ultra-subwavelength open end surface in the wave propagation direction to achieve ultra-subwavelength focusing. Moreover, we perform wave propagation simulations by using the software COMSOL. As a result, we find that the electric field is enhanced dozens of times, but most importantly, for the first time, we present an extremely enhanced magnetic field with an amplitude enhanced by five orders of magnitude. This approach is useful for obtaining extremely strong magnetic fields both for scientific study and for practical applications.

\section{One-way waveguide based on remanence}
We first study a straight waveguide which is a symmetric waveguide consisting of two YIG layers with opposed magnetization directions and separated by a dielectric layer. In this paper the dielectric is assumed to be glass with the (relative) permittivity $\epsilon_r = 2.25$\cite{Mason:30}. As shown in the inset of Fig. 1(a), the waveguide is covered in the y-direction by metal layers, which can be treated as perfect matched conductor (PEC) at microwave frequencies. In this Metal-YIG-Dielectric-YIG-Metal (MYDYM) model, because of the ferromagnetic properties, YIG layers still posses gyromagnetic properties after removing the external dc magnetic field and the (relative) permeability of YIG in the xyz basis takes the form
\begin{align}
    \overline\mu=\begin{bmatrix}\mu&-ik&0\\ik&\mu&0\\0&0&1\end{bmatrix},
\end{align}
where $\mu=1$, $k=-\omega_r/\omega$ ($\omega$ is the angular frequency). The parameters $\mu_0$, $\gamma$, $M_r$ and $\omega_r=\mu_0\gamma M_r$ are the vacuum permeability, the gyromagnetic ratio, the remanent magnetization and the characteristic circular frequency, respectively. In the proposed structure, the SMPs are transverse-electric-polarized (TE)\cite{Hartstein:31}, and based on the distribution of the electric field, they can be divided into two kinds of EM modes, i.e., even-symmetry (ES) and odd-symmetry (OS) modes. Derived from Maxwell's equations and the boundary conditions, the dispersion relations of the ES and OS modes can be respectively written as
\begin{equation}
    \mu_2k_x+\frac{\alpha_2}{{\rm tanh}(\alpha_2d_2)}+\alpha_1\mu_v{\rm tanh}(\alpha_1d_1)=0,\;\;\;\;\;\;\;\;\;(\rm even \ mode)\tag{2a}
\end{equation}

\begin{equation}
    \mu_2k_x+\frac{\alpha_2}{{\rm tanh}(\alpha_2d_2)}+\frac{\alpha_1\mu_v}{{\rm tanh}(\alpha_1d_1)}=0,\;\;\;\;\;\;\;\;\;\;\;\;(\rm odd \ mode)\tag{2b}
\end{equation}
where $d_2$ and $d_1$ represent the thicknesses of the YIG and dielectric layer. The variable $k_x$ is the propagation constant, $\alpha_1=\sqrt{k_x^2-\epsilon_rk_0^2}$ (where $k_0=\omega/c$, c is the speed of light in vacuum) and $\alpha_2=\sqrt{k_x^2-\epsilon_m \mu_v k_0^2}$ (where $\epsilon_m$ and $\mu_v = \mu-k^2/\mu$ are the permittivity and the Voigt permeability of YIG, respectively). From Eqs. (2a,b), one can easily calculate the asymptotic frequencies of the ES and OS modes for the two opposite propagation directions, i.e.,
\begin{equation}
    \omega_{\rm sp}^- = 0.5 \omega_r,\tag{3a}
\end{equation}
for $k_x\rightarrow - \infty$ and
\begin{equation}
    \omega_{\rm sp}^+ = \omega_r,\tag{3b}
\end{equation}
for $k_x\rightarrow+\infty$. It should be noted that $\omega_{\rm sp}^-$ and $\omega_{\rm sp}^+$ are the same for both the ES and OS modes. For convenience, we introduce a waveguide parameter $D=(d_1,d_2)$ (which is actually a vector of two parameters). In Fig. 1(a), the solid and dash-dotted lines respectively represent the dispersion curves of the ES and OS modes as $D=(0.2 \lambda_r,0.2 \lambda_r)$ (where $\lambda_r$ equals $2\pi c /\omega_r$). Throughout this paper, the characteristic remanent circular frequency and the permittivity of YIG are assumed to be $\omega_r = 2 \pi\times3.587\times10^9$ rad/s and $\epsilon_m = 15$\cite{Pozar:32}, respectively. The
\begin{figure}[ht]
\centering\includegraphics[width=4.5 in]{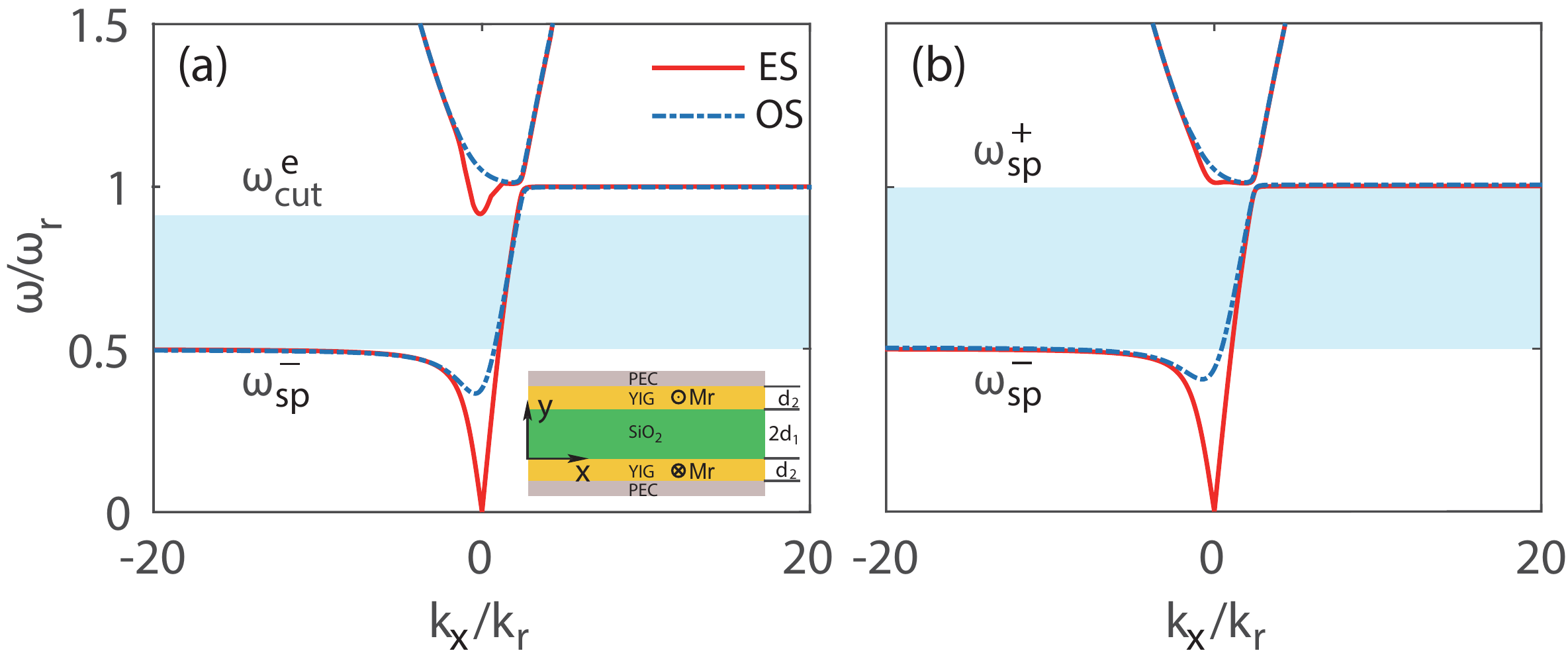}
\caption{ The solid and dash-dotted lines respectively represent the dispersion curves of the ES and OS modes as (a) $d_1=d_2=0.2\lambda_r$, (b) $d_1=0.16 \lambda_r$ and $d_2=0.2\lambda_r$. The shaded areas are the COWP regions. Inset: the schematic of the MYDYM structure. The other parameters are $\epsilon_r = 2.25$, $\epsilon_m=15$ and $\omega_r = 2 \pi \times 3.587 \times 10^{9}$ rad/s.}\label{Fig1}
\end{figure}
shaded area in Fig. 1(a) represents the complement one-way propagation (COWP) region and it obviously suffers from the high-frequency branch (the higher branch shown in Fig. 1(a)) of the ES modes. Although it is difficult to directly obtain the cutoff frequency ($\omega_{\rm cut}^{\rm e}$) of the high-frequency branch of the ES modes by solving Eq. (2a), the frequency $\omega_{\rm cut}^{\rm e}$ does not deviate much from the solution of Eq. (2a) in the limit $k_x \rightarrow 0$. When considering the asymptotic frequency (AF) band $(\omega_{\rm sp}^-$, $\omega_{\rm sp}^+)$, we have $\alpha_2 > 0$ for $\mu_v < 0$, and $\alpha_1 = i k_1$ with $k_1 = \sqrt{\epsilon_r}k_0$ as $k_x=0$ and Eq. (2a) changes into
\begin{equation}
    \frac{\alpha_2}{{\rm tanh}(\alpha_2 d_2)}-k_1\mu_v{\rm tan}(k_1 d_1)=0.\tag{4}
\end{equation}
In Eq. (4), the value of the first term is always greater than zero. Therefore, Eq. (4) has no solution as $k_1 d_1 < \pi/2$ in the AF band. Here, we introduce a critical thickness $d_c$ as $k_1 d_1 = \pi/2$ and $\omega_{\rm cut}^{\rm e}$ should be larger than $\omega_{\rm sp}^+$ when $d_1<d_c$. The critical thickness $d_c$ can be rewritten in the form
\begin{equation}
    d_c = \frac{1}{4 \sqrt{\epsilon_r}} \lambda_r.\tag{5}
\end{equation}
 In this paper, $d_c \approx 0.166 \lambda_r$ (for $\lambda_r \approx 83.6$mm) and Fig. 1(b) shows the dispersion curves of the ES and OS modes as $D= (0.16\lambda_r, 0.2 \lambda_r)$ in which $d_1$ is smaller than $d_c$. As a result, in Fig. 1(b) the COWP band is the whole AF band and we emphasize that the width of the COWP band in this case is maximal.

Furthermore, one can see from Fig. 1 that the low-frequency branch of the dispersion curves of the OS modes has a relative small cutoff frequency $\omega_{\rm cut}^{\rm o}$ $(\omega_{\rm cut}^{\rm o}<\omega_{\rm sp}^-)$. We further investigate the relation between the dispersion curves of the OS modes and the waveguide parameter D in Fig. 2(a), in which two horizontal dashed lines represent the asymptotic frequencies of the OS modes. Figs. 2(b) and (c) show the group velocity ($v_g$) of the OS modes as functions of $k_x$ and $\omega$. It is clear that for smaller $d_1$, e.g., for $d_1 = 0.01 \lambda_r$, $v_g$ is continuously larger than zero and $\omega_{\rm cut}^{\rm o}=\omega_{\rm sp}^-$. By contrast, in the cases $D_1$ and $D_2$, $\omega_{\rm cut}^{\rm o} < \omega_{\rm sp}^-$ and $v_g$ becomes negative in some frequencies below the AF band. Moreover, $v_g$ (in the AF band) in the $D_1$/$D_2$ cases is larger than for the $D_3$/$D_4$ cases, as shown in Fig. 2(c). On the other hand, comparing the cases with same $d_1$ value, i.e., $D_1$/$D_3$ and $D_2$/$D_4$, $v_g$ increases in the AF band when $d_2$ changes from $0.01\lambda_r$ to $0.1\lambda_r$. The ability of changing group velocity by simply changing the thickness of the YIG or the dielectric can be used to slow down or accelerate the propagation of EM modes. However, in this paper we just concentrate on the subwavelength focusing.
\begin{figure}[pt]
\centering\includegraphics[width=3 in]{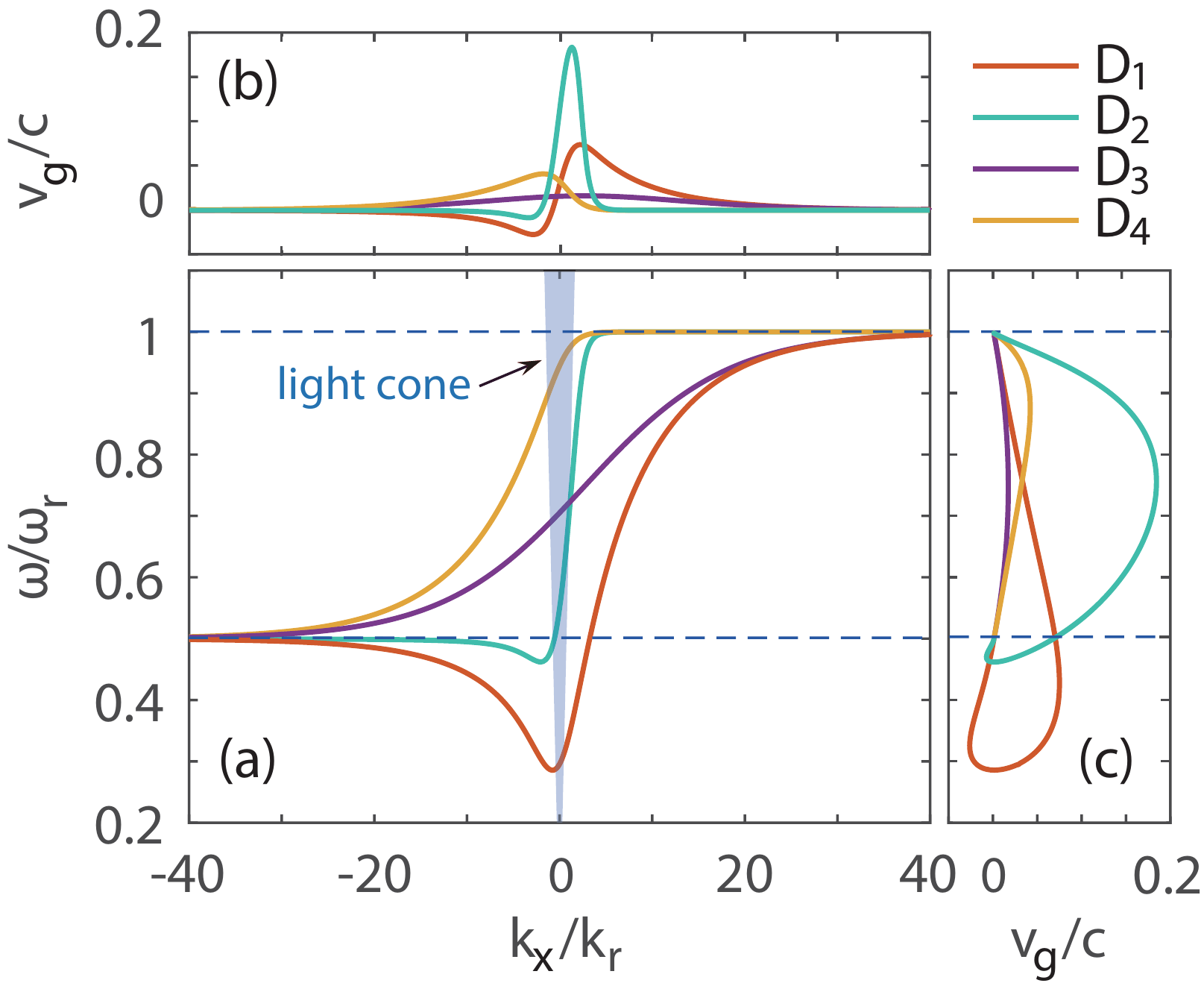}
\caption{(a) The dispersion curves and (b, c) the group velocities of the OS modes. The solid lines from bottom to top represent $D_1= \left (0.1\lambda_r, 0.01 \lambda_r\right)$, $D_2= \left (0.1\lambda_r, 0.1 \lambda_r\right)$, $D_3= \left (0.01\lambda_r, 0.01 \lambda_r\right)$ and $D_4= \left (0.01\lambda_r, 0.1 \lambda_r\right)$, respectively. The two horizontal lines represent the two asymptotic frequencies $\omega_{\rm sp}^{\pm}$ of Eq. (3) of the OS modes. The shaded zone is the light cone of the glass. The other parameters are the same as in Fig. 1.}\label{Fig2}
\end{figure}

Since it seems that $\omega_{\rm cut}^{\rm o}=\omega_{\rm sp}^-$ holds for cases with small values of $d_1$ in Fig. 2, we plot Fig. 3 that shows the functional relation between $\omega_{\rm cut}^{\rm o}$ and $d_1$ for different values of $d_2$. In accordance with the above discussion of Fig. 2, in Fig. 3(a) when $d_2$ increases, $\omega_{\rm cut}^{\rm o}$ increases and when $d_1$ increases, then $\omega_{\rm cut}^{\rm o}$ decreases, except for small $d_1$ cases in which $\omega_{\rm cut}^{\rm o} = \omega_{\rm sp}^-$. We further demonstrate the numerical values of $d_1$ and $d_2$ as $\omega_{\rm cut}^{\rm o} = \omega_{\rm sp}^-$ as the red line shown in Fig. 3(b). For clarity, in Fig. 3(b) we shade the left  and right areas of the red line, and the green and gray areas represent the OS modes with $\omega_{\rm cut}^{\rm o} = \omega_{\rm sp}^-$ and $\omega_{\rm cut}^{\rm o} \leq \omega_{\rm sp}^-$. We emphasize that for large $d_1$ cases, for example $d_1=0.1\lambda_r$, $\omega_{\rm cut}^{\rm o}\leq \omega_{\rm sp}^-$ permanently holds for all values of $d_2$. However, in this paper, we only consider the small $d_1$ cases ($d_1<0.01\lambda_r$) for designing ultra-subwavelength focusing waveguide.
\begin{figure}[pt]
\centering\includegraphics[width=4.5 in]{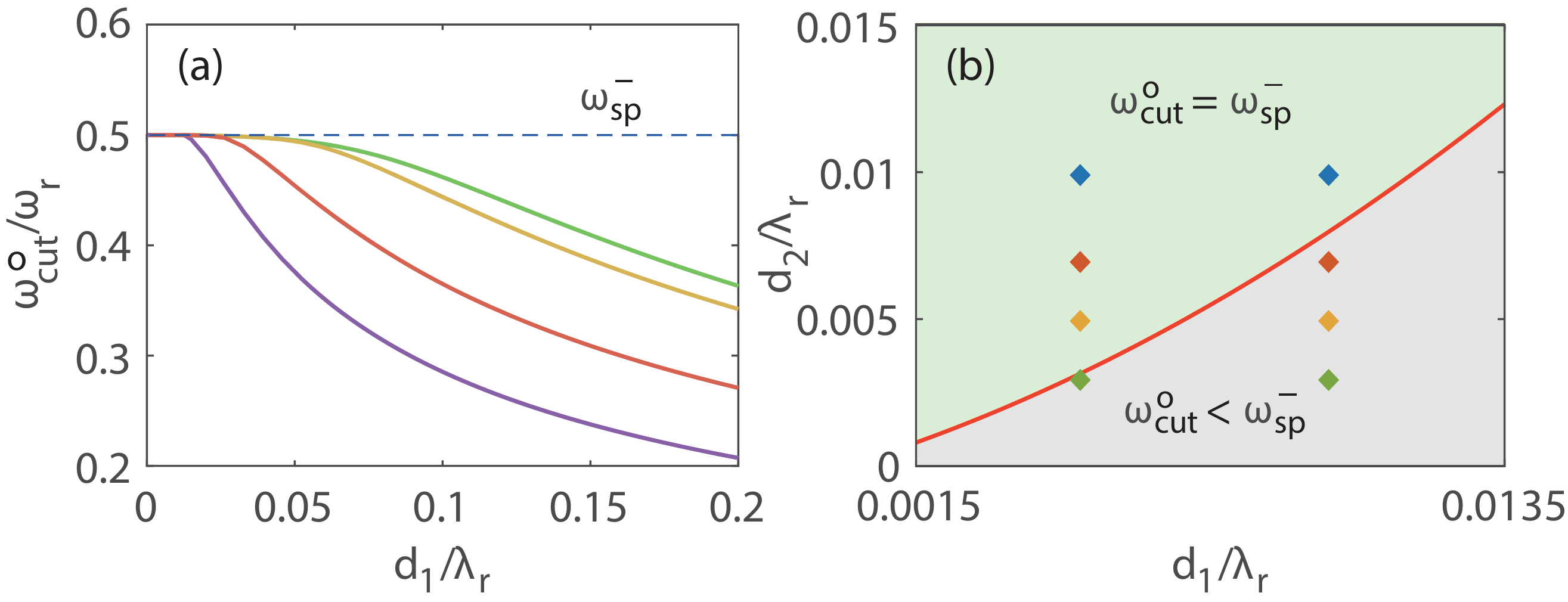}
\caption{ (a) The functional dependence of $\omega_{\rm cut}^{\rm o}$ on $d_1$ as $d_2$ is kept constant. Purple line: $d_2=0.01\lambda_r$; Red line: $d_2=0.02\lambda_r$; Yellow line: $d_2=0.05\lambda_r$; Green line: $d_2=0.1\lambda_r$. (b) The ($d_1$, $d_2$) space divided into two areas, i.e., the green shaded area and the gray shaded area and they represent $\omega_{\rm cut}^{\rm o} = \omega_{\rm sp}^-$ and $\omega_{\rm cut}^{\rm o} < \omega_{\rm sp}^-$, respectively.}\label{Fig3}
\end{figure}

\section{Ultra-subwavelength focusing waveguide}
\begin{figure}[ht]
\centering\includegraphics[width=4.5 in]{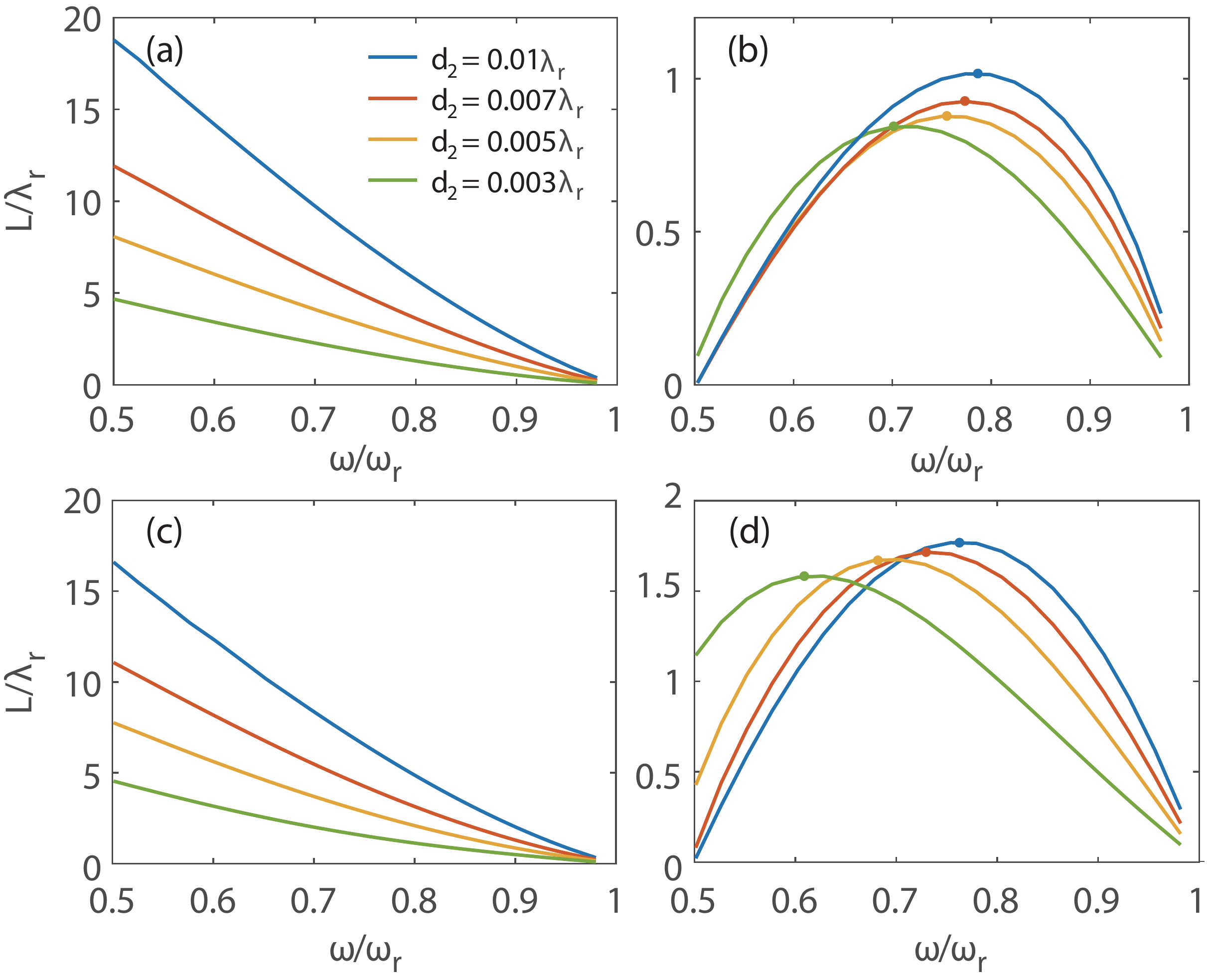}
\caption{ (a) (c) The L of the ES modes in the band ($\omega_{\rm sp}^-$, $\omega_{\rm sp}^+$) with $d_1 = 0.005\lambda_r$ in (a) and $d_1=0.01\lambda_r$ in (c). (b) (d) The L of the OS modes with $d_1 = 0.005\lambda_r$ in (b) and $d_1=0.01\lambda_r$ in (d). The relaxation angular frequency $v=10^{-3}\omega$.}\label{Fig4}
\end{figure}

\begin{figure}[ht]
\centering\includegraphics[width=4.5 in]{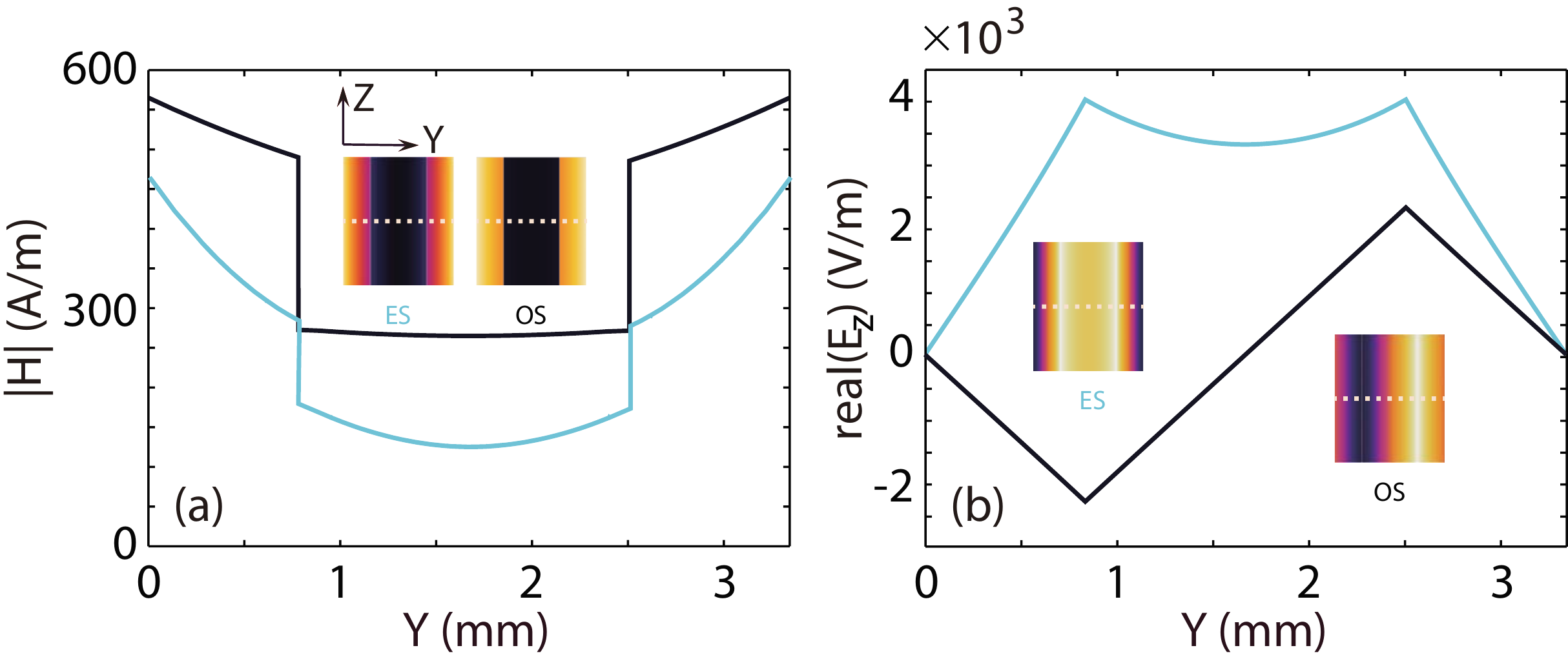}
\caption{ (a) The magnetic-field amplitude and (b) the real part of the electric-field amplitude of the ES modes (cyan lines) and the OS mode (black lines) along the dotted lines shown in the insets. The other parameters are $D=(0.01\lambda_r,0.01\lambda_r)$ and $\omega=0.75\omega_r$.}\label{Fig5}
\end{figure}
\begin{figure}[ht]
\centering\includegraphics[width=4.5 in]{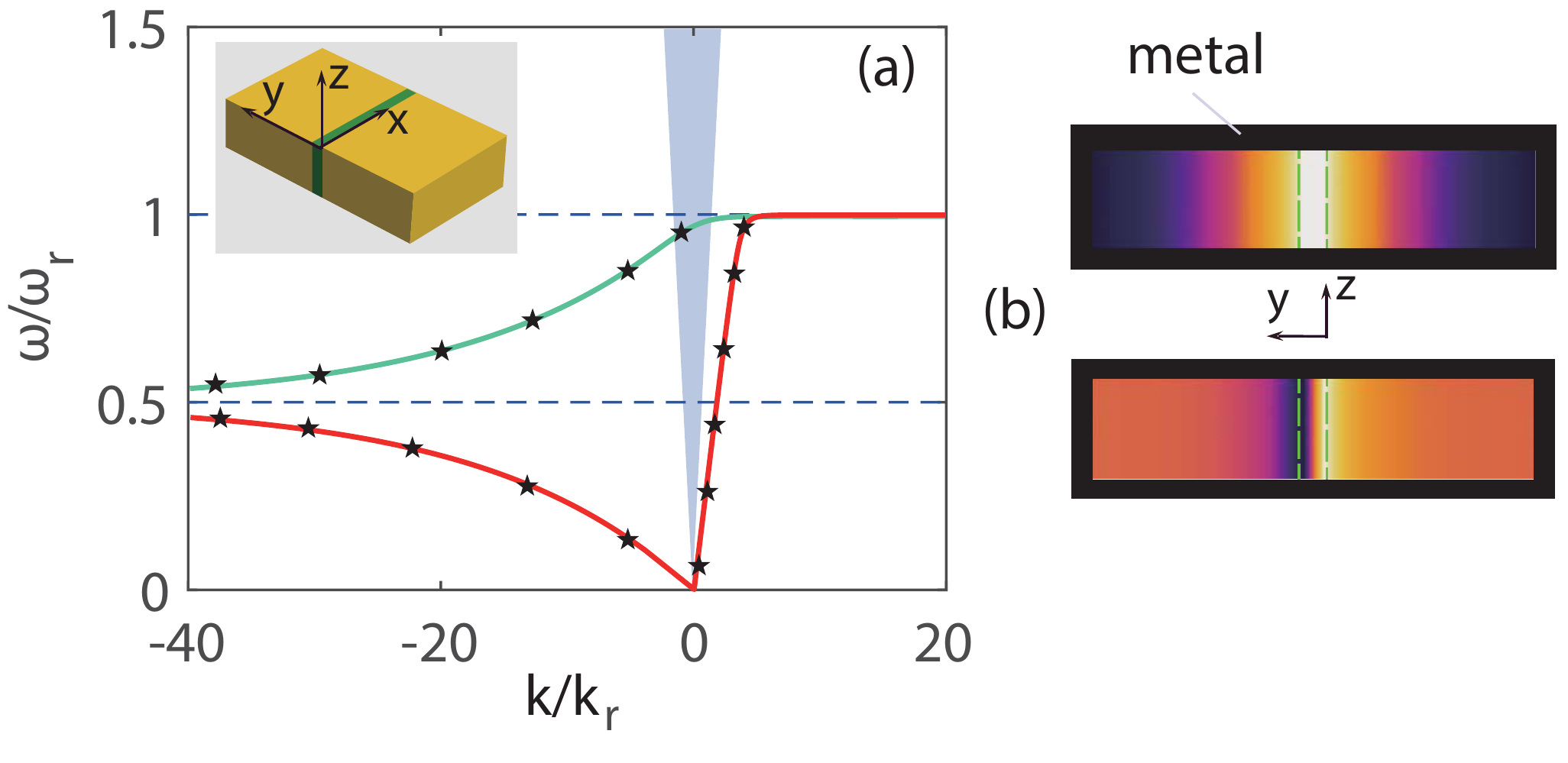}
\caption{ (a) The solid lines represent the dispersion curves of a 2D MYDYM waveguide as $d_1 = 0.005\lambda_r$ and $d_2=0.1\lambda_r$, and the stars are the eigen-mode solutions of the corresponding 3D model. The inset is the schematic structure of the 3D model and the height is set to be $h_1 = 4$ mm. (b) The electric-field ($E_z$) distribution of the ES (top) and OS (bottom) eigenmodes in the 3D model. The other parameters are same as in Fig. 1.}\label{Fig6}
\end{figure}
\begin{figure}[ht]
\centering\includegraphics[width=4.5 in]{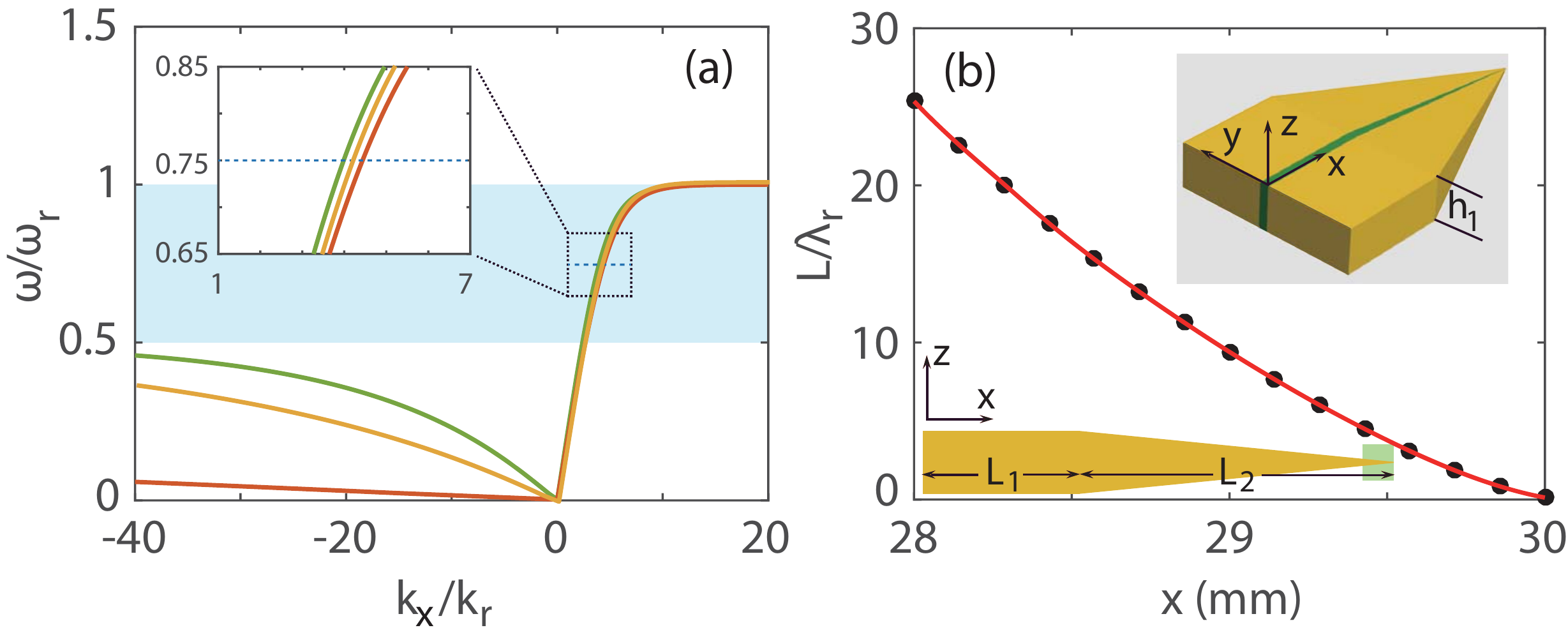}
\caption{ (a) The dispersion curves of the ES modes as $x=10$mm (the green lines), $x=20$mm (the yellow lines) and $x=30$mm (the red lines). (b) The L of the 3D model as $\omega = 0.75\omega_r$ as $28<x<30$ mm, i.e., the shaded green area shown in the left inset. The two insets show the schematic of the designed tapered 3D structure which consists of a straight part with length $L_1=10$ mm and a tapered part with length $L_2=20$ mm. The relaxation angular frequency is set to be $v=10^{-3}\omega$.}\label{Fig7}
\end{figure}

In the above analysis, we ignored the loss in the waveguide which should be a good approximation in calculating the dispersion relation\cite{Shen:33,Shen:34,You:15}. However, when designing a waveguide, loss should be carefully studied. In the lossy YIG material (see chapter 9 in \cite{Pozar:32}), $k$ and $\mu$ in Eq. (1) become
\begin{align}
    \mu = 1+ \frac{iv\omega_r}{\omega^2+v^2},\tag{6}
\end{align}
\begin{align}
    k = - \frac{\omega \omega_r}{\omega^2+v^2},\tag{7}
\end{align}
where $v$ is the relaxation angular frequency (in this paper, we set $v = 10^{-3}\omega$). In this lossy model, the propagation constant in Eqs. (2) and (3) now becomes a complex number, i.e., $k_x = k_r + i k_i$. The propagation length (L) is defined by $L = 1/(2 k_i)$. Fig. 4 shows the L as a function of the angular frequency $\omega$ for different waveguide parameters D. Four colored lines in Fig. 4 represent four different values of $d_2$ as the insert legend shows in Fig. 4(a). Figs. 4(a, c) and (b, d) represent the L of the ES and OS modes, respectively. Moreover, Figs. 4(a, b) and (c, d) have different values of $d_1$, i.e., $d_1 = 0.005\lambda_r$ in panels (a,b) and $d_1 = 0.01 \lambda_r$ in panels (c,d). As seen in Fig. 4(a,c), for the ES modes, the lower the frequency or the thicker the YIG layer, the longer propagation length. Fig. 4(b, d) shows that for the OS modes, the longest L happens at $0.5\omega_r < \omega <\omega_r$ (the marked points) and a "redshift" occurs as $d_2$ decreases. In Fig. 4(b, d), when $\omega \rightarrow 0.5\omega_r (\omega_{\rm sp}^-)$ the L in some cases vanishes, which can be understood when we look back to Fig. 3(b). In Fig. 3(b) the left/right four diamonds represent the four parameters in Fig. 4(b)/(d) and for those diamonds located in the green area with $\omega_{\rm cut}^{\rm o} = \omega_{\rm sp}^-$, we have $k_r \rightarrow \infty$, further, $k_i \rightarrow \infty$ and $L \rightarrow 0$, which indeed agree with the results shown in Fig. 4(b, d). Most importantly, comparing the four panels in Fig. 4, one can see that the L of the ES modes can be several or even dozens of times larger than the one of the OS modes. In other words, the ES modes propagate in the waveguide with much lower loss than the OS modes.

To further explain the low-loss propagation property of the ES modes, we solve the eigen-field distributions when the energy flux density is the same in the cross section (the YZ plane) for both the ES and OS modes in Fig. 5. The parameters are $D=(0.01\lambda_r,0.01\lambda_r)$ and $\omega=0.75\omega_r$ which is the center frequency of the AF band. The cyan and black lines in Fig. 5(a) respectively represent the magnetic-field amplitude $\left| H \right|$ of the ES and OS modes along the dotted lines shown in the inset. Fig. 5(b) is the corresponding real part of the electric field $E_z$ of the ES and OS modes. It is obvious in Fig. 5(a) that $\left| H \right|$ of the ES modes is smaller than the one of the OS modes in the YIG layers. In our calculation, about $54\%$ of the EM energy is confined in the YIG layers for the ES modes. By contrast, this value is nearly $94\%$ for the OS modes. Due to the gyromagnetic properties of YIG, the higher energy ratio in YIG will lead to a shorter propagation length. Therefore, we choose the ES modes as our preferred guided modes in designing the 3D subwavelength waveguide.

In Fig. 6(a) we compare the dispersion relation in straight uniform 2D and 3D models (see the inset) when the waveguide parameter D is the same in the two models. The red (cyan) line and stars respectively represent the dispersion curves of the ES (OS) modes solved in lossless 2D and 3D models. Note that in this paper, the 3D waveguides in the y- and z- directions are terminated by metal layers and for convenience, we only show the YIG-glass-YIG structure in all schematic structures of 3D models. We choose the waveguide parameter $D= \left (0.005\lambda_r, 0.1\lambda_r \right )$ because in the lossy 2D model with $v=10^{-3}\omega$, the propagation loss is extremely low and the propagation length exceeds $100\lambda_r$ as $\omega = 0.75\omega_r$. The propagation properties of 2D and 3D configurations are quite similar. This is because the EM modes in the proposed models are transverse electric (TE) modes and the non-zero electric field is $E_z$, which is perpendicular to the metal layers in the z-direction. Therefore, the EM mode distribution is almost the same for both 2D and 3D models. In Fig. 6(b), we show the electric ($E_z$) distribution of ES (top) and OS (bottom) eigenmodes in the cross section of the 3D model, and one can see that $E_z$ is uniform in the z-direction.

Inspired by the above analyses of Figs. 4, 5 and 6, we propose that in the COWP band ($\omega_{\rm sp}^-$, $\omega_{\rm sp}^+$), low-loss one-way 3D waveguides can be realized, and we further use this waveguide to achieve subwavelength focusing. In previous subwavelength focusing works \cite{He:21,Shen:10}, the focusing scale is always being $\sim 10^{-2} \lambda_0$ ( $\lambda_0$ is the wavelength in vacuum), and ultra-subwavelength ($\sim 10^{-4}\lambda_0$) focusing has never been achieved. Here, we propose a joint straight-tapered waveguide as the insets shown in Fig. 7(b). The parameters of the straight part are $D=(0.005\lambda_r,0.1\lambda_r)$, $L_1 = 10$ mm (the length of the straight part) and $h_1=4$ mm (the height of the straight part). The end surface of the tapered part has the parameters $d_1=d_2=2 \times 10^{-4}\lambda_r$ ($\sim 10^{-4}\lambda_0$, $\lambda_0 \approx 111.5$ mm as $\omega = 0.75\omega_r$) and $h_2  \approx 1.8 \times 10^{-4}\lambda_r$ ($\sim 10^{-4}\lambda_0$). The length of the tapered part is $L_2 = 20$ mm. Fig. 7(a) shows the dispersion curves of the ES modes in the tapered part as $x=10$mm (the green lines), $x=20$mm (the yellow lines) and $x=30$mm (the red lines). One can easily see from Fig. 7(a) that the ES modes propagating in the tapered part still possess the one-way propagation properties in the whole AF band. To further demonstrate the propagation loss in the waveguide, we numerically calculate the L in the region $28$ mm $\leq x \leq 30$ mm corresponding to the green shaded area in the left inset in Fig. 7(b) for $\omega=0.75 \omega_r$ (see the horizonal lines in Fig. 7(a)). As a result, we find a propagation length $L \approx 800$ mm as $x = 29$ mm. That is only 1 mm away from the end surface and even at the end surface ($x=30$ mm) $L$ still has a value of around 11 mm. In this designed structure, because of one-way low-loss propagation, one can expect significantly enhanced fields in the near-field region of the end surface.

\section{Extremely enhanced magnetic field}
\begin{figure}[ht]
\centering\includegraphics[width=4 in]{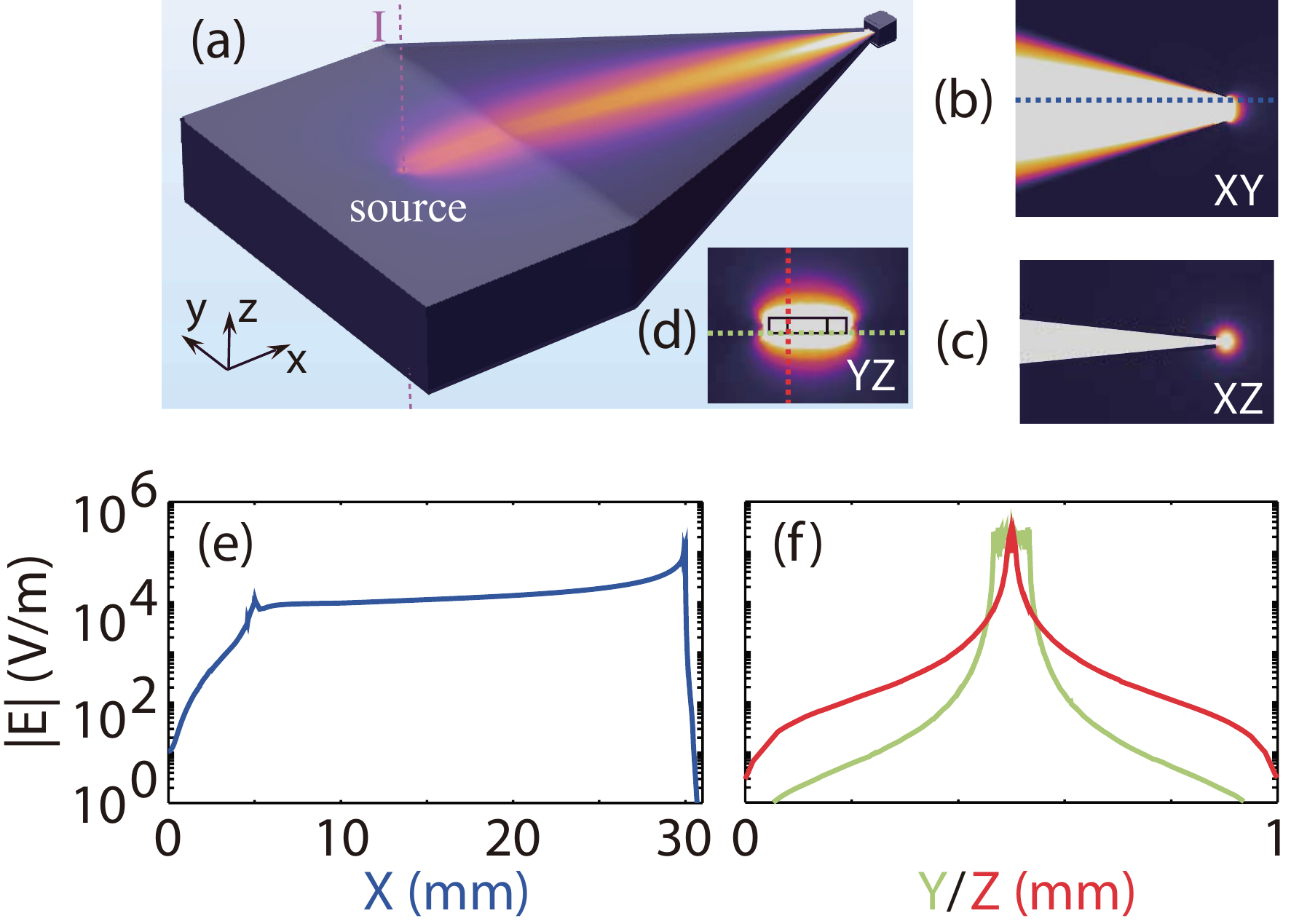}
\caption{ (a) The electric-field amplitude $\left| E \right|$ distribution in the designed 3D waveguide. The zoom magnified pictures of (a) near the end surface of the waveguide in (b) the XY plane ($z=-h_1/2$) and (c) the XZ plane ($y=d_1$). (d) The electric-field distribution on the end surface ($x=30$ mm). (e) The eletric-field amplitude along the dashed line shown in (b). (f) The eletric-field amplitude along the dashed lines shown in (d).}\label{Fig8}
\end{figure}
\begin{figure}[ht]
\centering\includegraphics[width=4 in]{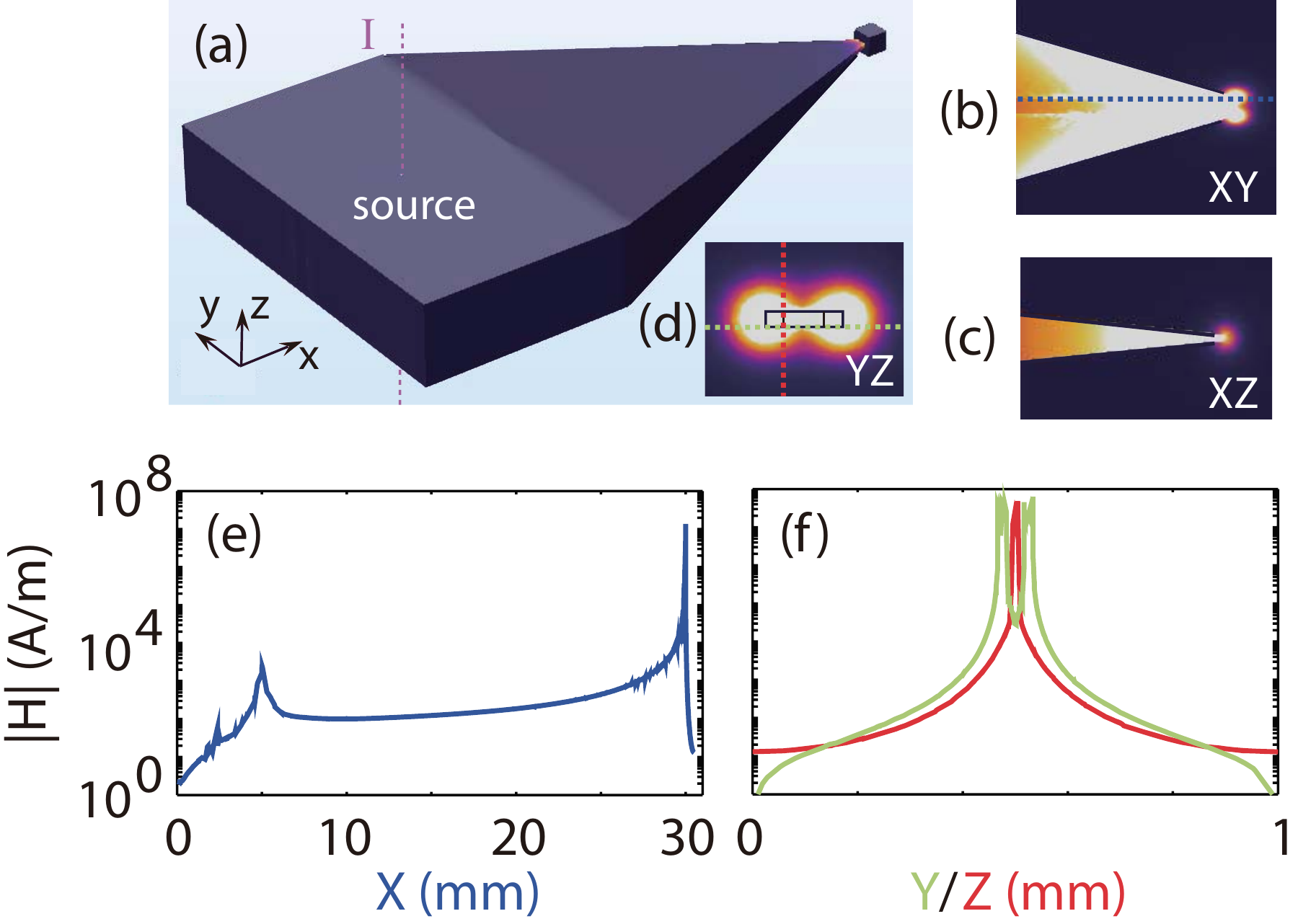}
\caption{ Same as in Fig. 8 except for the field shown in this figure is the magnetic field. (e) and (f) show that the magnetic field around the end surface is enhanced by five orders of magnitude compared to the straight waveguide.}\label{Fig9}
\end{figure}
\begin{figure}[ht]
\centering\includegraphics[width=4.5 in]{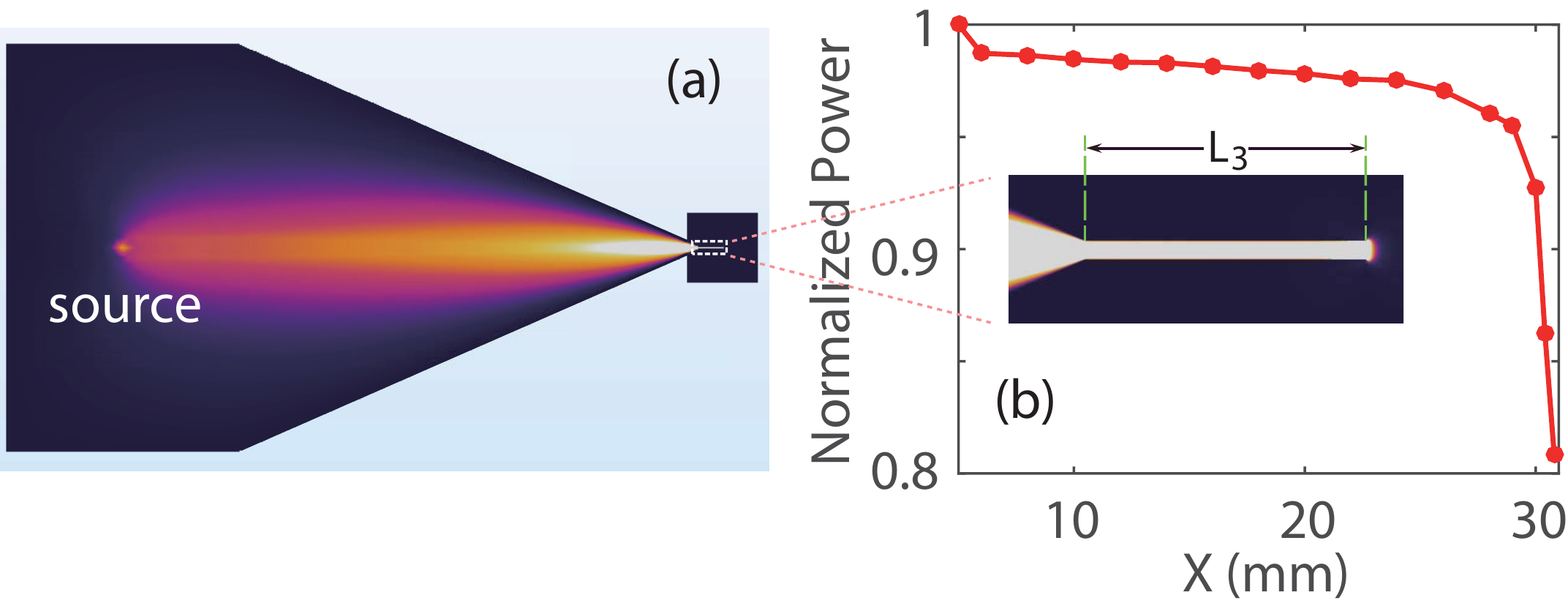}
\caption{ (a) The electric-field distribution in the lengthened 3D structure with a length $L_3=1$ mm added straight part. (b) The normalized power along the +x direction in the cross section of the model. The inset is the zoom-in picture of (a).}\label{Fig10}
\end{figure}
By using the commercial finite-element software COMSOL Multiphysics, we simulate the wave propagation in the above designed joint waveguide at $\omega = 0.75 \omega_r$ with $v=10^{-3}\omega$. As seen in Fig. 8(a), we set a line current source ($I=1$ A) in the middle of the glass layer to excite the ES modes and the coordinates of the source are $x = 5$ mm and $y=d_1$. Fig. 8(a) demonstrates the electric-field amplitude ($\left|E\right|$) distribution. As we expect, the excited mode propagates along the $+x$ direction with no interference even in the tapered waveguide part. Figs. 8(b), (c) and (d) show the zoom-in electric-field distributions in the XY ($z=-h_1/2$), XZ ($y=d_1$) and YZ (the end surface, $x=30$ mm) planes. It is quite clear that the electric field is focused in the near-field region around the end surface. Fig. 8(e) shows $\left|E\right|$ along the dashed line in Fig. 8(b) and we find that the electric field is enhanced dozens of times near $x=30$ mm compared with the value in the straight (i.e., non-tapered) waveguide. Similarly, in Fig. 8(f) the green and red lines respectively show the electric-field distribution along the green and red dashed lines marked in Fig. 8(d). Through Fig. 8(f), we can immediately calculate the full widths at half maximum (FWHM) of the focused electric field on the end surface to be FWHM$_y \approx 0.1$ mm ($ \approx 9 \times 10^{-4} \lambda_0$) and FWHM$_z$ $\approx 0.04$ mm  ($\approx 3.6 \times 10^{-4} \lambda_0$).

Fig. 9, being similar to Fig. 8, shows the corresponding simulation results for the magnetic field. In Fig. 9(a), the magnetic field is focused at the end of the waveguide. Figs. 9 (b), (c) and (d) are the magnetic-field distributions in the same planes corresponding to Figs. 8(b), (c) and (d). Because YIG is a ferromagnetic material, the magnetic field near the YIG layers will be larger than the one near glass or air as seen in Fig. 9(b) (the heart-shaped magnetic field distribution), Fig. 9(d) (the dumbbell-shaped magnetic field distribution) and Fig. 9(f) (the double peaks green line). Fig. 9(f) shows the distribution of the focused magnetic field on the end surface alone the dashed lines shown in Fig. 9(d). As one of our main results we shown in Fig. 9(e), that along the dashed line shown in Fig. 9(b), the magnetic-field amplitude is extremely enhanced by a few hundred thousand times near $x=30$ mm compared to the straight waveguide. As far as we know, this giant enhancement is never being achieved before.

In addition, we further perform a simulation to demonstrate the transmission efficiency and the loss in a new designed 3D model, in which we extrude the end surface of the previous model to build a straight waveguide with the length $L_3 = 1$ mm (see the inset in Fig. 10(b)). Fig. 10(a) shows the electric-field distribution in the XY plane ($z=-h_1/2$) and the inset in Fig. 10(b) is the zoom-in picture of Fig. 10(a). According to the results of Fig. 4 and Fig. 7, the EM mode in this configuration will propagate to the end with more and more loss, which can be seen in Fig. 10(b): the value of the slope of the normalized power flows gradually enlarged along $+x$ direction. However, the power efficiency remained about $93\%$ at $x=30$ mm, which implies the designed joint 3D model in Figs. 7, 8 and 9 is truly low-loss for the ES modes in the COWP band. Moreover, the propagation loss between $x=30$ mm and $x=31$ mm is around $10 \%$ which fits well with the results in Fig. 7(b). This approach to obtain extremely enhanced magnetic fields can be used in numerous fields including but not limited to magnetic field enhanced or quenched fluorescence, luminescence in 2D materials, novel optical tips for near-field microwave microscopy. Besides, because of the one-way propagation and ultra-subwavelength configuration, the nonlinearity around the end surface will also be increased (not considered here). Therefore this proposed one-way waveguide structure with giant magnetic-field enhancement has potential applications like energy storage.

\section{Conclusion and outlook}
In conclusion, we propose a tapered 3D MYDYM waveguide based on remanence to achieve ultra-subwavelength ($\sim10^{-4} \lambda_0 \times 10^{-4} \lambda_0$) focusing at microwave frequencies. We theoretically analyse the propagation properties of the ES and OS modes in the MYDYM structure, and compared to the OS modes, the ES modes propagate with much lower loss. When the thicknesses of the YIG layer and dielectric layer are identical, we found that the dispersion relation in the 3D model is almost the same as the one in the 2D model. By using the software COMSOL, we perform wave propagation simulations in the designed 3D model as $\omega=0.75\omega_r$ (the center frequency of the complement one-way propagation band). As we expected, the EM fields are confined in the tapered part and the electric field is enhanced dozens of times. More interestingly, in this case, the enhanced magnetic-field amplitude is enhanced up to five orders of magnitude. In addition, we further investigate the propagation loss in an extended tapered waveguide. The ES mode propagating in this waveguide is proved to be low-loss and the efficiency of the power in the simulation is about $93\%$ at the $x=30$ mm (the end surface of the previous waveguide). It should be noted that the mechanism mentioned in this paper can be expanded to the terahertz regime\cite{xu:9} once the equivalent perfect-magnetic-conductor (PMC) is built, for example by using metamaterials.

\section*{Funding information}
We acknowledge support by National Natural Science Foundation of China (NSFC)
(61372005), National Natural Science Foundation of China (NSFC) under a key project (41331070), and Independent Research Fund Denmark (9041-00333B). The Center for Nanostructured Graphene is sponsored by the Danish National Research Foundation (Project No. DNRF103).




\end{document}